\documentclass[reprint,aps,prb,showpacs,superscriptaddress]{revtex4-1}
\usepackage{amssymb,amsmath}
\usepackage{graphicx}
\usepackage{dcolumn}
\usepackage{multirow}
\usepackage{color}
\usepackage[english]{babel}
\usepackage{hyperref}
\usepackage{cases}
\usepackage{fixltx2e}
\usepackage{bm}
\usepackage{enumitem}

\newcommand{\bra}[1]{ \langle #1 | }
\newcommand{\ket}[1]{ |{#1} \rangle }

\newcommand*{\citen}[1]{%
	\begingroup
	\romannumeral-`\x 
	\setcitestyle{numbers}%
	\cite{#1}%
	\endgroup   
}

\begin{document}

\def\simlt{\mathrel{\lower .3ex \rlap{$\sim$}\raise .5ex \hbox{$<$}}}
\def\simgt{\mathrel{\lower .3ex \rlap{$\sim$}\raise .5ex \hbox{$>$}}}

\title{\textbf{\fontfamily{phv}\selectfont Adiabatic two-qubit gates in capacitively coupled quantum dot hybrid qubits 
}}
\author{Adam Frees}
\email{frees@wisc.edu}
\affiliation{Department of Physics, University of Wisconsin-Madison, Madison, WI 53706}
\author{Sebastian Mehl}
\affiliation{JARA-Institute for Quantum Information, RWTH Aachen University, D-52056 Aachen, Germany}
\affiliation{Peter Gr{\"u}nberg Institute (PGI-2), Forschungszentrum J{\"u}lich, D-52425 J{\"u}lich, Germany}
\author{John King Gamble}
\affiliation{Center for Computing Research, Sandia National Laboratories, Albuquerque, NM 87123}
\affiliation{Quantum Architectures and Computation Group, Microsoft Research, Redmond, WA 98052}
\author{Mark Friesen}
\affiliation{Department of Physics, University of Wisconsin-Madison, Madison, WI 53706}
\author{S. N. Coppersmith}
\affiliation{Department of Physics, University of Wisconsin-Madison, Madison, WI 53706}

\begin{abstract}

The ability to tune qubits to flat points in their energy dispersions (``sweet spots") is an important tool for mitigating the effects of charge noise and dephasing in solid-state devices. 
However, the number of derivatives that must be simultaneously set to zero grows exponentially with the number of coupled qubits, making the task untenable for as few as two qubits.
This is a particular problem for adiabatic gates, due to their slower speeds.
Here, we propose an adiabatic two-qubit gate for quantum dot hybrid qubits, based on the tunable, electrostatic coupling between distinct charge configurations. 
We confirm the absence of a conventional sweet spot, but show that controlled-Z (CZ) gates can nonetheless be optimized to have fidelities of $\sim$99\% for a typical level of quasistatic charge noise ($\sigma_\varepsilon$$\simeq$1~$\mu$eV).
We then develop the concept of a dynamical sweet spot (DSS), for which the time-averaged energy derivatives are set to zero, and identify a simple pulse sequence that achieves an approximate DSS for a CZ gate, with a 5$\times$ improvement in the fidelity.
We observe that the results depend on the number of tunable parameters in the pulse sequence, and speculate that a more elaborate sequence could potentially attain a true DSS.

\end{abstract}

\maketitle

{\bf Introduction ---}
 Since their original proposal,\cite{Loss:1998p120} semiconductor quantum dot qubits have progressed greatly, demonstrating excellent qubit coherence and performance through the use of
 sweet spots \cite{PhysRevLett.105.246804,PhysRevLett.110.146804,Kim:2014p70,KimNatNano15,PhysRevLett.116.086801,PhysRevLett.116.110402,PhysRevLett.116.116801,Schoenfield:2017aa,Thorgrimsson:2017aa,Mi:2018aa,Samkharadze1123} and
 control of the spin degree of freedom. \cite{Kawakami11738,Yoneda:2018aa,PhysRevLett.120.137702,2017arXiv170704357J}
 There has also been remarkable progress in systems with small numbers of donor-bound electrons.
  \cite{Pla:2012p541,Fuechsle:2012p242,Pla:2013p334,PhysRevLett.115.166806,Laucht:2016aa,Watsone1602811,Tosi:2017aa,Harvey-Collard:2017aa,Broome:2018aa}
Recently, two-qubit gates \cite{Veldhorst:2015aa,Zajac439} and algorithms \cite{Watson:2018aa} have been realized using exchange-coupled single-spin qubits.  
Capacitive coupling has also been employed to entangle and perform two-qubit operations between singlet-triplet qubits,\cite{Shulman202, Nichol:2017aa}
and has been proposed as the basis for two-qubit gates between resonant-exchange qubits \cite{PhysRevLett.111.050502} and flip-flop qubits \cite{Tosi:2017aa}.
In these experiments and proposals, the two-qubit gate times are typically measured in microseconds or hundreds of nanoseconds, which is much longer than typical single-qubit gate times.
In conrast, the predicted two-qubit gate times for capacitively-coupled quantum dot hybrid qubits\cite{2015arXiv150703425M,Ferraro2017} (QDHQs) are comparable to single-qubit gates, which are of order 10~ns.\cite{Kim:2015aa,Thorgrimsson:2017aa}
However, the methods proposed in refs~\citen{2015arXiv150703425M} and \citen{Ferraro2017} rely on applying quickly varying electrical pulses, which can cause leakage from the qubit subspace. \cite{Kim:2015aa}

In this paper we study an adiabatic entangling protocol based on capacitive couplings between QDHQs. 
The gate is inspired by an early proposal for entangling singlet-triplet qubits. \cite{Taylor:2005aa} 
Although the necessary voltage changes are slow relative to the qubit frequencies, we show that high-fidelity adiabatic gates can be achieved in under 50 ns, which is significantly faster than those in recent singlet-triplet experiments. \cite{Shulman202, Nichol:2017aa} 
While the pulse sequences used in adiabatic protocols are more resilient against pulse errors than non-adiabatic pulses
and are less susceptible to leakage errors, a potential concern is that they could be more susceptible to charge noise due to slower speeds. 
It is therefore crucial to study the effect of charge noise on the gate fidelities.

We begin by considering the system of two capacitively coupled QDHQs, deriving the effective couplings between the two qubits, and describing how a slowly-varying electrical pulse on the qubits can yield an entangling gate. Next, we optimize the pulse sequence for a two-qubit system to maximize the process fidelity of the resulting gate. We find that gate fidelities $>99\%$ are feasible, assuming quasistatic charge noise with a standard deviation of $\sigma_\varepsilon = 1$ $\mu$eV, and that the infidelity scales roughly as $\sigma_\varepsilon^2$. Finally, we show that gate fidelities can be further improved to $\sim$99.9\% by modifying pulse sequences to impose a ``dynamical sweet spot," a technique similar to dynamical decoupling.\cite{PhysRevLett.82.2417}

\section*{}
\begin{figure}[tb]
\includegraphics[width=1.0 \linewidth]{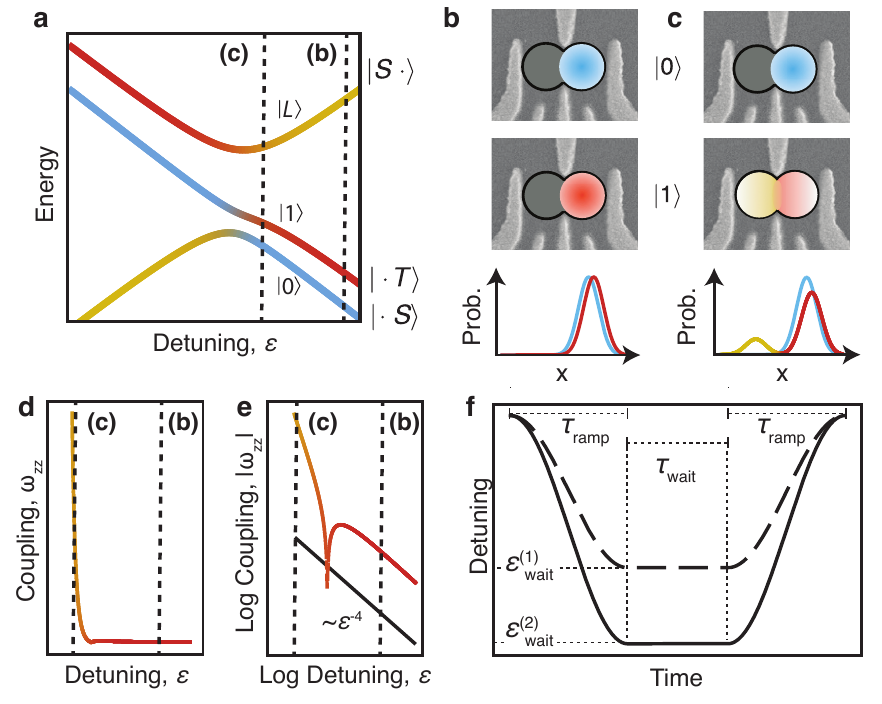} 
\caption{{\bf Implementing a CZ gate between capacitively coupled QDHQs.} 
{\bf a}, Energy dispersion of a single QDHQ as a function of detuning, as defined in Eq.~(\ref{single_H}), for typical experimental values given by\cite{Thorgrimsson:2017aa} $\Delta_1=18.1$ $\mu$eV, $\Delta_2=46.7$ $\mu$eV, and $E_\text{ST}=51.7$ $\mu$eV.
On the right-hand side of the plot, the logical states $\ket{0}$ and $\ket{1}$ converge to the basis states $\ket{\cdot S}$ (indicated in blue) and $\ket{\cdot T}$ (red), as defined in the main text, while the leakage state $\ket{L}$ converges to $\ket{S \cdot}$ (yellow).
{\bf b,c}, Charge distributions of a third electron added to an underlying (1,1) charge configuration at two different values of $\varepsilon$. 
{\bf b}, In the large-$\varepsilon$ regime (right-most dashed line in {\bf a}), the qubit states have very similar charge distributions (same color coding as {\bf a}).
{\bf c}, For $\varepsilon$ near the charge transition (left-most dashed line in {\bf a}), some charge moves from the right dot to the left dot, especially for state $\ket{1}$, setting up a dipole moment between states $\ket{0}$ and $\ket{1}$.
{\bf d,e}, Effective two-qubit coupling versus $\varepsilon=\varepsilon^{(1)}=\varepsilon^{(2)},$ plotted on linear-linear ({\bf d}) and log-log ({\bf e}) scales. 
When $\varepsilon$ is large, the coupling is negligible and decreases as $\varepsilon^{-4}$ (see Eq.~\eqref{coupling}). 
When $\varepsilon$ is simultaneously lowered on both qubits, their dipole moments grow, and the effective coupling increases.
{\bf f}, Detuning pulse sequences for qubits 1 and 2 (dashed and solid lines, respectively; see Supplementary Section~S3 for details).
\label{fig1}}
\end{figure}

{\bf Results ---}
The QDHQ consists of three electrons shared between two quantum dots. 
The minimal Hilbert space of the qubit can be defined as the spin states 
$\ket{\cdot S} =\ket{\downarrow} \ket{S}$, $\ket{\cdot T} = \sqrt{\frac{1}{3}}\ket{\downarrow} \ket{T_0} -\sqrt{\frac{2}{3}}\ket{\uparrow} \ket{T_-}$, and $\ket{S\cdot} = \ket{S}\ket{\downarrow}$,\cite{PhysRevLett.108.140503,PhysRevLett.109.250503} where $\ket{\cdot S}$ and $\ket{\cdot T}$ correspond to (1,2) charge configurations (one electron in the left dot, two electrons in the right), $\ket{S \cdot}$ corresponds to a (2,1) charge configuration, and the singlet state, $\ket{S}=1/\sqrt{2}\left(\ket{\downarrow\uparrow}-\ket{\uparrow\downarrow}\right)$, and triplet states, $\ket{T_0}=1/\sqrt{2}\left(\ket{\downarrow\uparrow}+\ket{\uparrow\downarrow}\right)$ and $\ket{T_-}=\ket{\downarrow\downarrow}$, refer to the dot with two electrons.
In this basis, the single-qubit Hamiltonian is 
\begin{equation}
\mathcal{H}_{1q}= \left(\begin{array}{ccc}
-\varepsilon /2 & 0 & \Delta_1 \\
0 & -\varepsilon /2 + E_\text{ST} & \Delta_2 \\
\Delta_1& \Delta_2& \varepsilon /2
\end{array}\right),\label{single_H}
\end{equation}
where the detuning parameter, $\varepsilon$, corresponds to the energy separation between the quantum dots,  $\Delta_1$ is the tunnel coupling between states $\ket{\cdot S}$ and $\ket{S \cdot}$, $\Delta_2$ is the tunnel coupling between states $\ket{\cdot T}$ and $\ket{S \cdot}$, and $E_\text{ST}$ is the energy splitting between the singlet-like and triplet-like basis states, $\ket{\cdot S}$ and $\ket{\cdot T}$.
A typical energy spectrum for $\mathcal{H}_{1q}$ is plotted as a function of detuning in Fig.~\ref{fig1}a.
Here, the two lowest-energy eigenstates $\ket{0}$ and $\ket{1}$ form the qubit, while the remaining state $\ket{L}$ is regarded as a leakage state.

The qubit states and charge configurations hybridize as a function of the detuning.
The large detuning regime (right-most dashed line in Fig.~\ref{fig1}a) is characterized by the asymptotic behavior $\ket{0}\simeq\ket{\cdot S}$ and $\ket{1}\simeq\ket{\cdot T}$, for which both states have the same charge configuration, as depicted in Fig.~\ref{fig1}b. 
Here, the information is stored entirely in the spin degree of freedom and the qubit is well protected from charge noise;\cite{Thorgrimsson:2017aa}
however the single-qubit gate speeds can  be slow.\cite{Wong:2016,Yang:2017} 
(Below, we show the same is true for two-qubit gates.)
To perform efficient gates, we must therefore lower the detuning, bringing it near the anticrossing region (left-most dashed line in Fig.~\ref{fig1}a).
In this regime, $\ket{\cdot S}$ and $\ket{\cdot T}$ begin to hybridize with $\ket{S\cdot}$, which has a (2,1) charge configuration, as depicted in Fig.~\ref{fig1}c.
Since the admixture of (2,1) is different for $\ket{0}$ and $\ket{1}$, the qubit states acquire distinct dipole moments that can be used to mediate two-qubit dipole-dipole interactions, but which also couple to environemental charge noise.
The goal of this work is to optimize the control parameters, to achieve high-fidelity two-qubit gates.

\begin{figure}[!tp]
	\includegraphics[width=1.0 \linewidth]{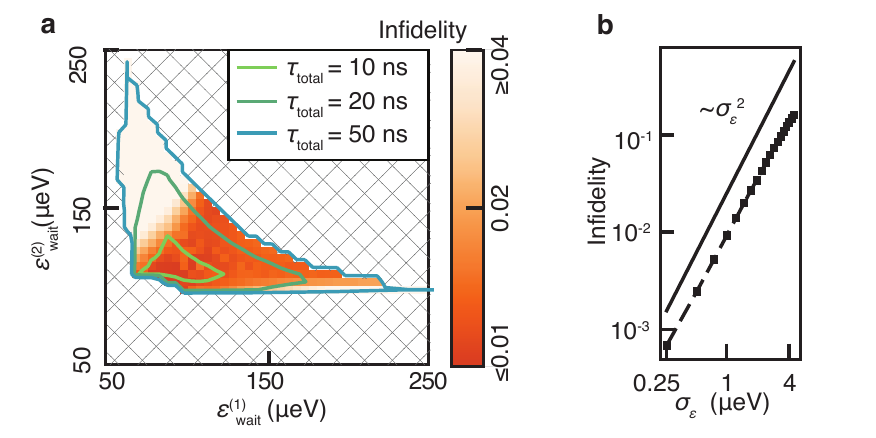} 
	\caption{{\bf Optimized infidelities of sub-50~ns adiabatic CZ gates.} 
{\bf a}, Process infidelities obtained in the presence of quasistatic charge noise, with a standard deviation of $\sigma_\varepsilon$=1~$\mu$eV; other Hamiltonian parameters are given in the main text. 
Following the optimization procedure described in Methods, we obtain variable gate times, as indicated by the contours; however, we discard total gate times with $\tau_{\text{total}}$=$2\tau_{\text{ramp}}$+$\tau_{\text{wait}}$$>50$~ns (cross-hatched region) to ensure that the two-qubit gate is comparable in length to single-qubit gates.
The highest fidelity pulse sequence is obtained at $(\varepsilon^{(1)}_{\text{wait}},\varepsilon^{(2)}_{\text{wait}})$=(80,100)~$\mu$eV. 
{\bf b}, Minimum CZ gate infidelities (black squares and dashed black line) plotted as a function of the  standard deviation of the charge noise, $\sigma_\varepsilon$. 
The infidelity roughly falls off as $\sigma_\varepsilon^2$ (solid black line).\label{fig2}}
\end{figure}

We can formalize the concept of a dipole moment by defining the operator $\hat{x}=\text{diag}\{d/2,d/2,-d/2\}$, describing the position of the third electron electron in the double dot, as depicted in Fig.~\ref{fig2}b,c.
Here for simplicity, we assume that states $\ket{\cdot S}$ and $\ket{\cdot T}$ have identical charge configurations.
The dimensionless dipole operator is therefore given by $\mathcal{P}=-\hat{x}/d=\text{diag}\{-1/2,-1/2,1/2\}$, which is related to Eq.~(\ref{single_H}) through $\mathcal{P} = \partial \mathcal{H}_{1q}/\partial \varepsilon$, where $\varepsilon$ plays the role of an electric field along the axis between the dots.
The two-qubit Coulomb interaction can be expressed in terms of the dipole moments $\mathcal{P}^{(1)}$ and $\mathcal{P}^{(2)}$ of qubits 1 and 2.
We first note that the Coulomb interaction is classical, and therefore diagonal, when expressed in a charge-state basis.
In analogy with charge qubits, the interaction can then be fully specified by $b^{(0)}\mathcal{I}^{(1)}\otimes\mathcal{I}^{(2)}+b^{(1)}\mathcal{P}^{(1)}\otimes\mathcal{I}^{(2)}+b^{(2)}\mathcal{I}^{(1)}\otimes\mathcal{P}^{(2)}+b^{(3)}\mathcal{P}^{(1)}\otimes\mathcal{P}^{(2)}$.
The first term in this expression is a uniform energy shift, which can be ignored.
The second and third terms can be absorbed into the detuning parameters through the transformation $\varepsilon^{(i)}\rightarrow\varepsilon^{(i)}+b^{(i)}$ ($i=1,2$).
Finally, defining $g$ as the change in Coulomb energy when one of the qubits flips its charge configuration, the two-qubit Hamiltonian becomes
\begin{equation}
\mathcal{H}_{2q} = \mathcal{H}_{1q} ^{(1)}\otimes I^{(2)} + I^{(1)} \otimes \mathcal{H}_{1q} ^{(2)} + g\, \mathcal{P}^{(1)} \otimes \mathcal{P}^{(2)} . \label{two_qubit}
\end{equation}
This form is generic and does not depend on qubit geometry.
However, the value of $g$ depends on the geometry, and has been found to be of order 75~$\mu$eV for a linear dot array.\cite{Ward:2016aa}
The full 9D basis set for Eq.~(\ref{two_qubit}) is given by $\{\ket{\cdot S},\ket{\cdot T},\ket{S\cdot}\}^{(1)}\otimes\{\ket{\cdot S},\ket{\cdot T},\ket{S\cdot}\}^{(2)}$, and the corresponding matrix representation for $\mathcal{H}_{2q}$ is given in Supplementary Section~S1.


We first discuss qubit initialization and the implementation of single-qubit gates.
In the large-detuning regime, the qubit logical states are energetically well separated from the leakage states, as shown in Supplementary Fig.~S1, allowing leakage-free qubit initialization.
To gain insight into gate operations, we perform a canonical transformation to decouple the logical states from the leakage states in the large-detuning limit.
Additionally, we evaluate this Hamiltonian in the adiabatic basis (the basis that diagonalizes $\mathcal{H}_{2q}$), yielding the effective Hamiltonian
\begin{multline}
\mathcal{H}_{2q,\text{eff}}\simeq 
\frac{-\hbar\omega_{z1}}{2} \sigma_z^{(1)}\otimes I^{(2)}  
+ \frac{-\hbar\omega_{z2}}{2} I^{(1)}\otimes\sigma_z^{(2)} \\
+ \frac{\hbar\omega_{zz}}{2}\sigma_z^{(1)}\otimes\sigma_z^{(2)} ,
\label{H2q_approx}
\end{multline}
where the leading-order contributions to the single-qubit prefactors are of order
$\hbar\omega_{zi}=E_\text{ST}^{(i)}+\mathcal{O}[{\Delta_z^{(i)}}^2/\varepsilon^{(i)}]$, and the effective two-qubit coupling $\hbar\omega_{zz}$ is discussed below.
(See Supplementary Section~S1 for details of the calculation.)
Here, the sub- (or super)-script $i$ identifies the qubit,  ${\Delta_z^{(i)}}^2$ is a quadratic function of the tunnel couplings, and the identity and Pauli matrices, $I^{(i)}$ and $\sigma_z^{(i)}$, act on the logical subspace. 
Although the detuning parameters provide some control over the qubit resonant frequencies, $\omega_{z1}$ and $\omega_{z2}$, $E_\text{ST}$ typically varies significantly from dot to dot, resulting in well separated resonances.
Single-qubit gates thus proceed by lowering one of the detunings (say, $\varepsilon^{(1)}$) from its high value to a regime where fast ac gates can be performed (e.g., the first dashed line in Fig.~\ref{fig1}a).
At this point, the dipole on qubit~1 is non-negligible; however, we can operate it near a single-qubit sweet spot to minimize dephasing, as described in Supplementary Section~2.
Since $\varepsilon^{(2)}$ remains at a large value, there is no danger of implementing either a single-qubit gate on qubit 2, or a two-qubit gate.
Superimposing an ac drive on $\varepsilon^{(1)}$ at the resonant frequency of qubit~1 yields an additional term in Eq.~(\ref{H2q_approx}) proportional to $\cos (\omega_{z1}t)\, \sigma_x^{(1)}\otimes I^{(2)}$, which induces Rabi oscillations about the $\hat{\bm{x}}$ axis of the qubit; additional modulation of the phase in $\cos (\omega_{z1}t+\phi)$ enables rotations about an arbitrary axis in the $x$-$y$ plane.
To suppress the coupling of the dipole moment to external charge noise, we return $\varepsilon^{(1)}$ to its large value when the gate is finished.

Next, we consider two-qubit gate operations, which are performed adiabatically, and do not involve ac driving.
The canonical transformation leading to Eq.~(\ref{H2q_approx}) yields the leading order result at high detuning, $\hbar\omega_{zz}=\mathcal{O}[g\Delta^4/{\varepsilon^{(1)}}^2{\varepsilon^{(2)}}^2]$, where $\Delta^4$ is a quartic function of the tunnel couplings in both qubits.
As anticipated, the adiabatic gate speed $|\omega_{zz}|/2\pi$ depends linearly on $g$, and requires $\varepsilon^{(1)}$ and $\varepsilon^{(2)}$ to be simultaneously reduced from their high values to initiate a two-qubit gate. 
The canonical transformation breaks down when $\varepsilon^{(1)}$ and $\varepsilon^{(2)}$ take their low values; however under adiabatic operation, the projection onto the logical subspace, Eq.~(\ref{H2q_approx}), is still meaningful.
We can compute $\hbar\omega_{zz}$ at arbitrary detuning values by evaluating $\mathcal{H}_{2q}$ in its adiabatic basis and projecting it onto the 4D logical subspace, 
$\mathcal{H}_{2q}\rightarrow\mathcal{H}_\text{4D}$.
We then identify $\hbar\omega_{zz} = \frac{1}{2}\text{Tr}[(\sigma_z^{(1)}\otimes\sigma_z^{(2)})\mathcal{H}_\text{4D}]=\frac{1}{2}(E_{00}-E_{01}-E_{10}+E_{11})$,
where $E_{ij}$ is the energy eigenvalue corresponding to the two-qubit logical state $\ket{ij}$.
In Fig.~\ref{fig1}d,e, we plot numerical results for $\hbar\omega_{zz}$ assuming typical qubit parameters and $\varepsilon\equiv\varepsilon^{(1)}=\varepsilon^{(2)}$.
Here we observe the predicted asymptotic behavior $\hbar\omega_{zz}\propto\varepsilon^{-4}$.
We also note that $\hbar\omega_{zz}$ changes sign when $\varepsilon$ is of order $g$, in the low-detuning regime where the canonical transformation breaks down.

A simple protocol for implementing adiabatic two-qubit gates is shown in Fig.~\ref{fig1}f, and can be summarized as follows. 
We begin with the detuning parameters $\varepsilon^{(1)}$  and $\varepsilon^{(2)}$ set to `high' values of 500~$\mu$eV, and smoothly lower them to the `low' values $\varepsilon^{(1)}_{\text{wait}}$ and $\varepsilon^{(2)}_{\text{wait}}$ over a ramp time $\tau_{\text{ramp}}$. 
The detunings are held constant at these values for a waiting period $\tau_{\text{wait}}$, and then smoothly returned to $\varepsilon^{(1)} = \varepsilon^{(2)} = 500$ $\mu$eV over the same ramp time $\tau_{\text{ramp}}$. 
The parameters defining the pulse sequence are chosen to approximately yield a controlled-Z (CZ) gate operation.
This protocol also produces incidental single-qubit $\text{Z}^{(1)}$ and $\text{Z}^{(2)}$ rotations, which can be eliminated, if necessary, by applying additional $\text{Z}^{(1)}$ and $\text{Z}^{(2)}$ gates.
Explicit functional forms for $\varepsilon^{(1)}(t)$  and  $\varepsilon^{(2)}(t)$ are given in Supplementary Section S3. 

We now compute the two-qubit gate fidelity for this sequence including both leakage and charge noise.
Leakage corresponds to the filling of quantum levels outside the logical subspace, and is primarily caused by non-adiabatic gate pulses.
It is taken into account in our simulations by retaining the full 9D Hilbert space, comprising both logical and leakage states.
In Methods we describe a method for computing the process fidelity of a CZ gate in the presence of charge noise. 
This procedure allows us to identify optimal values of $\tau_\text{ramp}$ and $\tau_\text{wait}$, consistent with fast pulse sequences, low leakage, and high fidelity.
Figure~\ref{fig2} shows the results of such fidelity calculations, for a range of $\varepsilon^{(1)}_{\text{wait}}$ and $\varepsilon^{(2)}_{\text{wait}}$ values, assuming the typical quantum dot parameters $E_\text{ST}^{(1)} = 52$~$\mu$eV, $E_\text{ST}^{(2)} = 47$~$\mu$eV, $g = 75$ $\mu$eV, and $\sigma_\varepsilon = 1$~$\mu$eV.
Here, we choose $\Delta_1^{(i)} = 0.64E_\text{ST}^{(i)}$ and $\Delta_2^{(i)} = 0.58E_\text{ST}^{(i)}$, which suppresses the single-qubit dephasing, as discussed in Supplementary Section~S2.
We also omit pulse sequences with total gate times $\tau_\text{total}=2\tau_\text{ramp}+\tau_\text{wait}>50$~ns (the cross-hatched regions in the plot), to ensure that entangling gates are performed on a timescale comparable to the QDHQ single-qubit gates.\cite{Kim:2015aa}
The fastest pulse sequence in the viable regime corresponds to $\varepsilon^{(1)}_{\text{wait}}= 90$~$\mu$eV, $\varepsilon^{(2)}_{\text{wait}} = 110$~$\mu$eV, $\tau_\text{ramp} = 2.4$~ns, and $\tau_\text{wait} = 2.8$~ns ($\tau_\text{total}=7.6$~ns), and exhibits an average process infidelity of $9.9 \times 10^{-3}$. 
The highest-fidelity sequence corresponds to $\varepsilon^{(1)}_{\text{wait}} = 80$~$\mu$eV, $\varepsilon^{(2)}_{\text{wait}} = 100$~$\mu$eV, $\tau_\text{ramp} = 4.0$~ns, and $\tau_\text{wait}= 8.0$~ns ($\tau_\text{total}=16.0$~ns), with an average process infidelity of $9.2 \times 10^{-3}$.
This optimized value depends on the standard deviation of the charge noise, $\sigma_\varepsilon$.
In Fig.~\ref{fig2}b, we plot the minimized CZ gate infidelity $\mathcal{I}$ as a function of $\sigma_\varepsilon$, revealing the scaling behavior $\mathcal{I}\propto\sigma_\varepsilon^2$. 

The strong dependence of infidelity on $\sigma_\varepsilon$ indicates that dephasing, rather than leakage, is the main source of gate errors.  
In Supplementary Section~S7, we explain the observed behavior by assuming that charge noise is quasistatic, obtaining 
\begin{equation}
\mathcal{I}_\text{cn} \approx \frac{1}{4} \sigma_\varepsilon^2\sum_{i = 1,2}\,\sum_{j = z1,z2,zz} \left( \int\frac{\partial \omega_j}{\partial \varepsilon^{(i)}} dt\right)^2 ,
\label{quadDerivatives}
\end{equation}
as expected in the absence of a sweet spot.\cite{PhysRevLett.116.086801}
To confirm the absence of a sweet spot, we perform an exhaustive search over the detuning ($\epsilon^{(i)}$), tunnel coupling ($\Delta_j^{(i)}$), and Coulomb ($g$) parameters in Eq.~(\ref{two_qubit}), finding that it is impossible to simultaneously set $\partial \omega_j/\partial \varepsilon^{(i)}=0$, for all $i$ and $j$, in the parameter range of interest.
However, it is clear that this conventional, time-independent definition of a sweet spot is overly restrictive for ensuring that $\mathcal{I}_\text{cn}\approx 0$ in Eq.~(\ref{quadDerivatives}).

 \begin{figure}[tb]
	\includegraphics[width=0.95 \linewidth]{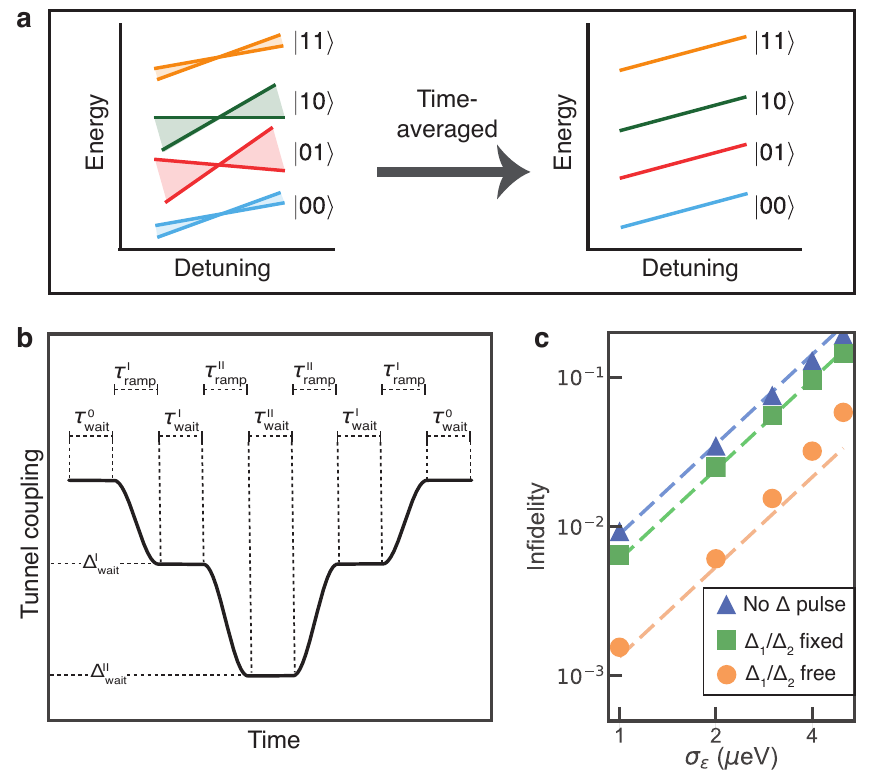} 
	\caption{{\bf Dynamical sweet spot (DSS) and a tunnel-coupling pulse sequence.}
{\bf a}, Two-qubit energy levels plotted schematically as a function of a single detuning variable. 
At any given time, it is difficult to arrange for all the energy dispersions to be parallel, as indicated on the left, leaving the qubits susceptible to dephasing.
However, it may be possible to construct a pulse sequence for which the levels vary in time (shaded regions), such that their time-averaged dispersions are parallel, yielding a DSS that is more resilient to quasistatic fluctuations of the detuning.
Here, we explore a DSS construction in which the detuning pulse sequence of Fig.~1f is augmented with the tunnel coupling pulse sequence defined in {\bf b} and Supplemental Section~S3.
The latter is simple enough that it can be optimized using the method described in the main text, but complex enough that it provides a significant improvement in the CZ gate fidelity.
{\bf c}, Infidelities calculated for three different pulse sequences as a function of the standard deviation of the quasistatic charge noise $\sigma_\varepsilon$. 
The markers correspond to full gate simulations averaged over a charge noise distribution, as described in Methods.
The dashed lines correspond to the much simpler infidelity estimate of Eq.~\eqref{quadDerivatives}.
For the blue line and triangles, the tunnel couplings are held constant, as in Fig.~\ref{fig1}.
For the green line and squares, the tunnel couplings are pulsed as in {\bf c} with the ratios $\Delta_1^{(i)}/\Delta_2^{(i)}$ held constant.
For the orange line and circles, the tunnel coupling sequence is optimized with no constraints.
\label{fig3}}
\end{figure}

We now introduce the concept of a dynamical sweet spot (DSS) in which the \emph{time-averaged} derivatives in Eq.~(\ref{quadDerivatives}) are made to vanish, as sketched in Fig.~3a. 
Through an exhaustive search (Fig.~2a), we have already demonstrated that no DSS exists for the simple pulse sequence of Fig.~1f.
Moreover, because of the monotonic dependence of $\omega_j$ on $\varepsilon^{(i)}$ (for example, see Supplementary Fig.~S1), it appears unlikely that a more elaborate detuning pulse sequence could provide significant improvements in the fidelity.
We therefore augment the detuning sequence with a tunnel-coupling sequence, $\Delta^{(i)}_j(t)$ ($i,j=1,2$).
Our initial investigations suggest that a relatively large number of pulse parameters are needed to achieve significant improvements in the fidelity.
We therefore consider the more elaborate pulse shape, shown in Fig.~\ref{fig3}b, which incorporates seven parameters for each of four tunnel couplings.

Because of the large number of parameters in the combined detuning-tunnel-coupling sequence, we do not attempt an exhaustive search in this case.
Instead, we maximize the CZ gate fidelity by performing a hundred separate Broyden-Fletcher-Goldfarb-Shanno \cite{doi:10.1093/imamat/6.1.76,fletcher1970new,goldfarb1970family,shanno1970conditioning} (BFGS) searches using the method of ref.~\citen{Jones:2001aa}, and adopting a basin-hopping protocol with randomized initial values to help escape any local minima.\cite{Leary2000}
To simplify the calculation, we adopt the following hybrid infidelity functional: $\mathcal{I}_\text{total}=\mathcal{I}_\text{cn}+\mathcal{I}_\text{nf}+\mathcal{I}_\text{na}$, which treats the charge noise (cn), noise-free (nf), and non-adiabatic (na) infidelity contributions separately.
Calculating $\mathcal{I}_\text{total}$ is computationally efficient because all three contributions, including the charge noise term defined in Eq.~(\ref{quadDerivatives}), do not require taking an average over charge noise.
The noise-free term describes the CZ gate infidelity in the absence of charge noise, as described in Methods.
We find that minimizing just the $\mathcal{I}_\text{cn}$ and $\mathcal{I}_\text{nf}$ terms (without $\mathcal{I}_\text{na}$) yields extremely short and fast pulse sequences that first populate then depopulate the leakage state. Since $\mathcal{I}_\text{cn}$ was derived assuming an adiabatic pulse, these short and fast pulses are not guaranteed to have low process infidelity.
Hence we also introduce the $\mathcal{I}_\text{na}$ term, as defined in Methods, to explicitly penalize non-adiabatic evolution.

We now obtain two different sets of solutions for the detuning-tunnel-coupling pulse sequence.
In the first, all the tunnel coupling parameters in Fig.~3b, as well as $\tau_\text{ramp}$ and $\tau_\text{wait}$, are varied independently, under the constraint that the detuning and tunnel coupling sequences have the same length; this sequence contains 26 free parameters.
The second case is similar, except that the ratio between the tunnel couplings in each double dot is assumed to be fixed throughout the sequence, with $\Delta^{(i)}_1(t)/\Delta^{(i)}_2(t)=1.1034$, as consistent with Supplementary Section~S2; this sequence contains 14 free parameters.
In both cases, we set the detuning parameters to the values giving the fastest detuning-only pulse sequence in Fig.~2 ($\varepsilon^{(1)}_\text{wait}=90$~$\mu$eV and $\varepsilon^{(2)}_\text{wait}=110$~$\mu$eV), and we use $\tau_\text{ramp} = 2.4$~ns and $\tau_\text{wait} = 2.8$~ns as the starting points for our optimization procedure; initial values of the other parameters are chosen randomly, according to the basin-hopping protocol.
The results of this procedure are presented in Supplementary Table~S1.
Using these results, we recompute the infidelity as described in Methods, performing and average over the charge noise.  

Infidelity results using the tunnel coupling pulse sequence are plotted in Fig.~3c as a function of charge noise, $\sigma_\varepsilon$.
Here we observe clear improvements compared to the detuning-only sequence, with the best results obtained for the sequence with the largest number of pulse parameters.
Supplementary Section~S8 suggests that this result can largely be attributed to the suppression of the time-averaged derivatives $\partial \omega_j/\partial \varepsilon^{(i)}$, as consistent with a DSS.
For a true sweet spot, we would expect a power-law exponent in $\mathcal{I}\propto\sigma_\varepsilon^\alpha$ , with $\alpha >2$.
Although the large-$\sigma_\varepsilon$ data in Fig.~3c hint at such behavior, Supplementary Figure~S6 indicates that a full suppression of the time-averaged derivatives has not yet been achieved in the current pulse sequences.

\section*{}
{\bf Discussion ---}
We have proposed a scheme for entangling capacitively coupled quantum-dot hybrid qubits by applying adiabatic pulse sequences to detuning parameters.  
We have optimized the sequences in the presence of quasistatic charge noise and computed the resulting process fidelities for a controlled-Z gate, obtaining fidelities approaching 99\% for typical noise levels. 
Further improvements are obtained by simultaneously applying pulse sequences to the tunnel couplings.
These results are explained by invoking the concept of a dynamical sweet spot (DSS), for which the splittings between the two-qubit energy levels are insensitive to fluctuations of the detuning parameters when averaged over the whole pulse sequence.
Our analysis shows that a true DSS cannot be achieved using simple pulse sequences.
However, fidelities $>$99\% are achieved when the pulse sequences include a large number of tunable parameters.
As indicated by ref \citen{Setser}, these fidelities can be further improved by exploring a wider range of pulse shapes.
We speculate that a bandwidth-limited version\cite{PhysRevA.97.062346,Frey:2017aa} of the GRAPE algorithm\cite{KHANEJA2005296} could be used to explore a much larger parameter space of adiabatic pulse sequences, possibly allowing us to identify a true DSS.
The GRAPE algorithm also provides a means for exploring non-adiabatic pulse sequences.
However the simplicity and relatively high fidelity achieved with the sequences studied here, and the robustness of adiabatic gating methods, make the current proposal attractive for two-qubit gates.

\section*{}
{\bf Methods ---}
To study the time evolution resulting from the pulse sequences applied to capitively coupled qubits, we numerically integrate the Hamiltonian in Eq.~(\ref{two_qubit}), for which the time-dependent control parameters $\tilde\varepsilon^{(1)}(t)$, $\tilde\varepsilon^{(2)}(t)$, and $\vec{\Delta}(t)$ depend on the particular pulse sequence.
Here, $\vec{\Delta}(t)$ refers to the set of four intra-qubit tunnel couplings, and we define $\tilde{\varepsilon}^{(i)}(t) = \varepsilon^{(i)}(t) + \delta\varepsilon^{(i)}$, where $\varepsilon^{(i)}(t)$ is the ideal, noise-free detuning sequence for qubit $i$, and the (quasi-static) noise term $\delta\varepsilon^{(i)}$ is assumed to remain constant for the duration of the sequence.
The resulting unitary operator is given by
 \begin{equation}
U_{2q}(t) = \exp\left[-i/\hbar \int_0^t \mathcal{H}_{2q}\left({\varepsilon}^{(1)}(t'),{\varepsilon}^{(2)}(t'),\vec{\Delta}(t')\right) dt' \right] .\label{unitary}
 \end{equation}
In most cases, we take $t$ to be the final time in the pulse sequence, with one exception, described below.

We employ the following procedure to determine the detuning pulse parameters used in Fig.~2.
(For additional details, see Section~S4 of the Supplementary Materials.)
We first choose the fastest ramp time $\tau_\text{ramp}$ consistent with leakage errors $<$0.1\% in the absence of charge noise. 
We then compute $U_{2q}$ as a function of $\tau_\text{wait}$ for a fixed level of quasi-static charge noise.
(High-frequency noise can also affect the fidelity of slow QDHQ gates;\cite{PhysRevB.86.035302} however we do not consider such processes here.)
The process fidelity $\mathcal{F}$ is computed, comparing $U_{2q}$ to a perfect CZ gate, modulo single-qubit rotations, using the Choi-Jamiolkowski formalism,\cite{PhysRevA.71.062310} as described in Supplementary Section~S5.
We then average $\mathcal{F}$ over charge noise, using the method described below, and choose the value of $\tau_\text{wait}$ that maximizes $\langle\mathcal{F}\rangle$.

To optimize the detuning-tunnel-coupling pulse sequence used in Fig.~3, we choose pulse parameters that minimize the total infidelity function $\mathcal{I}_\text{total}=\mathcal{I}_\text{cn}+\mathcal{I}_\text{nf}+\mathcal{I}_\text{na}$, as discussed in the main text.
Here, the noise-free term $\mathcal{I}_\text{nf}$ describes the CZ gate infidelity, computed using the Choi-Jamiolkowski formalism, as described above, in the absence of charge noise.
In this work, we also introduce a penalty term to suppress non-adiabatic evolution, defined as $\mathcal{I}_\text{na}=\text{max}_t\left[1-\frac{1}{4}\sum \left|\bra{ij(t)}U_{2q}(t)\ket{ij(0)}\right|^2\right]$,
where the sum is taken over the logical basis states $(i,j)=(0,1)$, and the function $\text{max}_t$ picks out the maximal occupation of leakage states, at any point in the pulse sequence. 
Note that the state $\ket{ij(0)}$ is an eigenstate of $H_{2q}(t)$ at time $t=0$, while $\ket{ij(t)}$ is the corresponding eigenstate at time $t$. 
Under perfect adiabatic operation, the mapping $U_{2q}(t)\ket{ij(0)}\rightarrow\ket{ij(t)}$ is exact, yielding $\mathcal{I}_\text{na}=0$; however for non-adiabatic operation, we obtain $\mathcal{I}_\text{na}>0$.
In practice, we find that the exact form of $\mathcal{I}_\text{na}$ does not significantly affect our results.

To average the fidelity over charge noise, we assume that the noise terms $\delta\varepsilon^{(1)}$ and $\delta\varepsilon^{(2)}$ are uncorrelated and sample them independently at 17 values in the range between -25 and +25~$\mu$eV, corresponding to 1089 unique pairs. 
We then linearly interpolate $\mathcal{F}$ over the results and calculate its average value, assuming a gaussian probability distribution with standard deviation $\sigma_\varepsilon$:
\begin{equation}
p(\delta\varepsilon^{(1)},\delta\varepsilon^{(2)}) = \frac{1}{2\pi\sigma_\varepsilon^2}\exp\left( - \frac{{\delta\varepsilon^{(1)}}^2 +{\delta\varepsilon^{(2)}}^2}
{2\sigma_\varepsilon^2}\right).
\end{equation}

\section*{}
{\bf Acknowledgements ---}
We thank D. Bradley, C. King, and R. Blume-Kohout for useful discussions. We also thank the HEP, Condor, and CHTC groups at University of Wisconsin-Madison for computational support.
This work was supported in part by ARO (W911NF-12-1-0607, W911NF-17-1-0274) and the Vannevar Bush Faculty Fellowship program sponsored by the Basic Research Office of the Assistant Secretary of Defense for Research and Engineering and funded by the Office of Naval Research through Grant No. N00014-15-1-0029. The views and conclusions contained in this document are those of the authors and should not be interpreted as representing the official policies, either expressed or implied, of the Army Research Office (ARO), or the U.S. Government. The U.S. Government is authorized to reproduce and distribute reprints for Government purposes notwithstanding any copyright notation herein.  This paper describes objective technical results and analysis. Any subjective views or opinions that might be expressed in the paper do not necessarily represent the views of the U.S. Department of Energy or the United States Government.  Sandia National Laboratories is a multimission laboratory managed and operated by National Technology \& Engineering Solutions of Sandia, LLC, a wholly owned subsidiary of Honeywell International Inc., for the U.S. Department of Energy’s National Nuclear Security Administration under contract DE-NA0003525.  The authors gratefully acknowledge support from the Sandia National Laboratories Truman Fellowship Program, which is funded by the Laboratory Directed Research and Development (LDRD) Program.

\section*{}
{\bf Author Contributions ---}
All authors contributed to the idea of using adiabatic gate sequences to entangle capacitively-coupled QDHQs. A.F. performed the Schrieffer-Wolff transformation. A.F. and J.K.G. performed the numerical simulation, and analyzed the data with M.F. and S.N.C. A.F., J.K.G., M.F., and S.N.C. wrote the manuscript and prepared the figures, with input from all the authors. 

\section*{}
{\bf Competing financial interests ---}
The authors declare that they have no competing financial interests.

\section*{}
{\bf Additional Information ---}
Supplementary information accompanies this paper. 
Correspondence and requests for materials should be addressed to Adam Frees (frees\emph{@}wisc.edu).

\clearpage
\setcounter{figure}{0}
\setcounter{equation}{0}
\renewcommand{\theequation}{S\arabic{equation}}
\renewcommand{\thefigure}{S\arabic{figure}}
\renewcommand{\thetable}{S\arabic{table}}
\begin{widetext}
\section*{Supplemental material for ``Adiabatic two-qubit gates in capacitively coupled quantum dot hybrid qubits"}
These supplemental materials provide additional details about the methods used in this work.

\section{DERIVATION OF AN EFFECTIVE HAMILTONIAN FOR TWO COUPLED QDHQS}

In this section, we derive an effective 4D Hamiltonian describing the logical subspace of two capacitively coupled QDHQs. 
In the basis $\{\ket{\cdot S},\ket{\cdot T},\ket{S\cdot}\}^{(1)}\otimes\{\ket{\cdot S},\ket{\cdot T},\ket{S\cdot}\}^{(2)}$, the full 9D Hamiltonian, Eq.~\eqref{two_qubit} of the main text, is given by
\begin{equation}
\mathcal{H}_{2q} = \left(
\begin{array}{cccc|cccc|c}
E_0 & 0 &   0 & 0 & \Delta_1^{(2)} & \Delta_1^{(1)} &   0 & 0 & 0 \\
0 & E_1 & 0& 0 & -  \Delta_2^{(2)}  & 0 & 0 &   \Delta_1^{(1)} & 0 \\
0 & 0 & E_2 & 0 & 0 & -  \Delta_2^{(1)} & \Delta_1^{(2)}  & 0 &   0\\
0 & 0 & 0 & E_3 & 0 &   0 & -  \Delta_2^{(2)} & -  \Delta_2^{(1)}& 0 \\
\hline
\Delta_1^{(2)} & -  \Delta_2^{(2)} & 0 & 0 & E_4 & 0 & 0 & 0 & \Delta_1^{(1)}\\
 \Delta_1^{(1)} & 0 & -  \Delta_2^{(1)} &   0 & 0 & E_5 & 0 & 0 &  \Delta_1^{(2)} \\
0& 0 & \Delta_1^{(2)}  & -  \Delta_2^{(2)} & 0 & 0 & E_6 & 0 &   -\Delta_2^{(1)} \\
0 &  \Delta_1^{(1)}  & 0 & -  \Delta_2^{(1)} & 0& 0 & 0 & E_7 & -  \Delta_2^{(2)} \\
\hline
0 & 0 &   0 & 0 & \Delta_1^{(1)}  &  \Delta_1^{(2)} &   -\Delta_2^{(1)} & -  \Delta_2^{(2)} & E_8 \\
\end{array}
\right),\label{fullHam}
\end{equation} 
where we adopt the same notation as the main text, and define
\begin{gather}
E_0 = -\frac{\varepsilon^{(1)}}{2}+\frac{g }{4}-\frac{\varepsilon^{(2)}}{2}, \\
E_1 = -\frac{\varepsilon^{(1)}}{2}+\frac{g }{4}+E_\text{ST}^{(2)}  -\frac{\varepsilon^{(2)}}{2}, \\
E_2 = -\frac{\varepsilon^{(1)}}{2}+\frac{g }{4}+E_\text{ST}^{(1)}  -\frac{\varepsilon^{(2)}}{2}, \\
E_3 = -\frac{\varepsilon^{(1)}}{2}+\frac{g }{4}+E_\text{ST}^{(1)}  +E_\text{ST}^{(2)}  -\frac{\varepsilon^{(2)}}{2}, \\
E_4 = -\frac{\varepsilon^{(1)}}{2}+\frac{\varepsilon^{(2)}}{2}-\frac{g }{4}, \\
E_5 = \frac{\varepsilon^{(1)}}{2}-\frac{\varepsilon^{(2)}}{2}-\frac{g }{4} , \\
E_6 = -\frac{\varepsilon^{(1)}}{2}+\frac{\varepsilon^{(2)}}{2}+E_\text{ST}^{(1)}  -\frac{g }{4} , \\
E_7 = \frac{\varepsilon^{(1)}}{2}+E_\text{ST}^{(2)}  -\frac{\varepsilon^{(2)}}{2}-\frac{g }{4} , \\
E_8 = \frac{\varepsilon^{(1)}}{2}+\frac{\varepsilon^{(2)}}{2}+\frac{g }{4}.
\end{gather}
In Eq.~(\ref{fullHam}), the solid lines delineate three distinct subspaces.
$E_0$ through $E_3$ represent the logical manifold, in which the energy levels decrease with $\varepsilon^{(1)}$ and $\varepsilon^{(2)}$).
$E_4$ through $E_7$  represent a leakage manifold in which $\varepsilon^{(1)}$ and $\varepsilon^{(2)}$ have opposite effects on the energy.
$E_8$ is an additional leakage state for which the energy increases with $\varepsilon^{(1)}$ or $\varepsilon^{(2)}$).
Some typical eigenvalues of Eq.~(\ref{fullHam}) are plotted in Fig.~\ref{figsupp0}, where we have set $\varepsilon^{(1)}=\varepsilon^{(2)}$ and added $\varepsilon^{(1)}$ to all the eigenstates.
At large detunings, we observe a large energy splitting between the logical and leakage manifolds.
To gain insight into the gate operations, we can therefore perform a Schrieffer-Wolff \cite{schrieffer-wolff} decomposition to adiabatically eliminate the leakage states.
Working to fourth order, and further diagonalizing the resulting 4D Hamiltonian, we obtain the leading-order contributions to the prefactors in Eq.~(\ref{H2q_approx}) of the main text:
\begin{gather}
\hbar\omega_{z1} = E_\text{ST}^{(1)} + \frac{{\Delta_2^{(1)}}^2-{\Delta_1^{(1)}}^2}{\varepsilon^{(1)}-g/2} + \mathcal{O}(\varepsilon^{-2}), \\
\hbar\omega_{z2} = E_\text{ST}^{(2)} + \frac{{\Delta_2^{(2)}}^2-{\Delta_1^{(2)}}^2}{\varepsilon^{(2)}-g/2} + \mathcal{O}(\varepsilon^{-2}), \\
\hbar\omega_{zz} = 
\frac{8g\left(g-\varepsilon^{(1)}-\varepsilon^{(2)}\right)
\left({\Delta_1^{(1)}}^2-{\Delta_2^{(1)}}^2\right) \left({\Delta_1^{(2)}}^2-{\Delta_2^{(2)}}^2\right)}{\left(g-2\varepsilon^{(1)}\right)^2\left(g-2\varepsilon^{(2)}\right)^2(\varepsilon^{(1)}+\varepsilon^{(2)})} .\label{coupling}
\end{gather}
\end{widetext}
Here, we have also assumed that $E_\text{ST}^{(i)}\ll\varepsilon^{(i)}-g/2$.
For additional details, please see the Supplemental Mathematica file.

\begin{figure}[ht]
	\includegraphics[width=1.0 \linewidth]{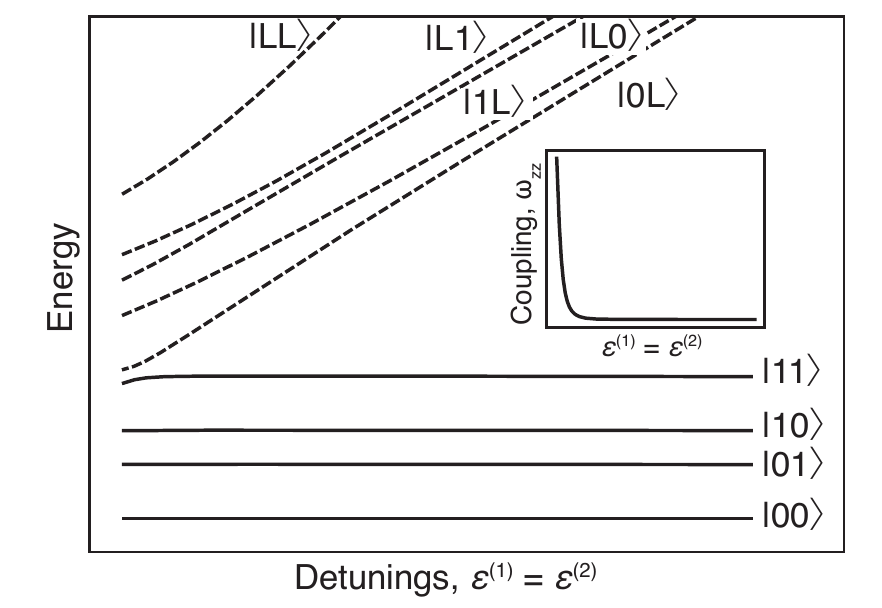} 
	\caption{{\bf Energy spectrum of two capacitively-coupled QDHQs as a function of qubit detunings.}
In the limit of large detuning, the low-energy logical subspace is well separated from the leakage states.
For smaller detuning values, the mixing between the leakage states and the low-energy states grows, giving rise to an effective Coulomb interaction between the qubits 
(inset -- inter-qubit coupling $\omega_{zz}$ of four-dimensional effective Hamiltonian, see Eq.~\ref{H2q_approx} of main text). 
The results shown here assume $E_\text{ST}^{(1)} = 52$ $\mu$eV, $E_\text{ST}^{(2)} = 32$ $\mu$eV, $\Delta_1^{(i)} = 0.64E_\text{ST}^{(i)}$, $\Delta_2^{(i)} = 0.58E_\text{ST}^{(i)}$, and $g=-75$ $\mu$eV. 
(See Section S6, below, for discussion about the sign of $g$.)
		 \label{figsupp0}}
\end{figure}
\section{Optimizing single qubit dispersions}

To achieve high-fidelity two-qubit gates, one must suppress both single- and two-qubit errors. Dephasing due to charge noise is a significant source of single-qubit errors in semiconductor-based quantum dot qubits, but can be mitigated by tuning the qubit near a sweet spot where the splitting between energy levels of the logical states, $\hbar \omega$, is insensitive to small fluctuations of the detuning, i.e., $\partial \omega / \partial \varepsilon = 0$.
Since the pure dephasing time $T^*_2$ is inversely proportional to $|\partial \omega / \partial \varepsilon|$,\cite{Thorgrimsson:2017aa,PhysRevLett.110.146804} it can increase significantly near such a sweet spot.
Here, we identify a working regime for a QDHQ where approximate sweet spot behavior can be achieved over a wide range of $\varepsilon$.

\begin{figure}[hb]
	\includegraphics[width=1.0 \linewidth]{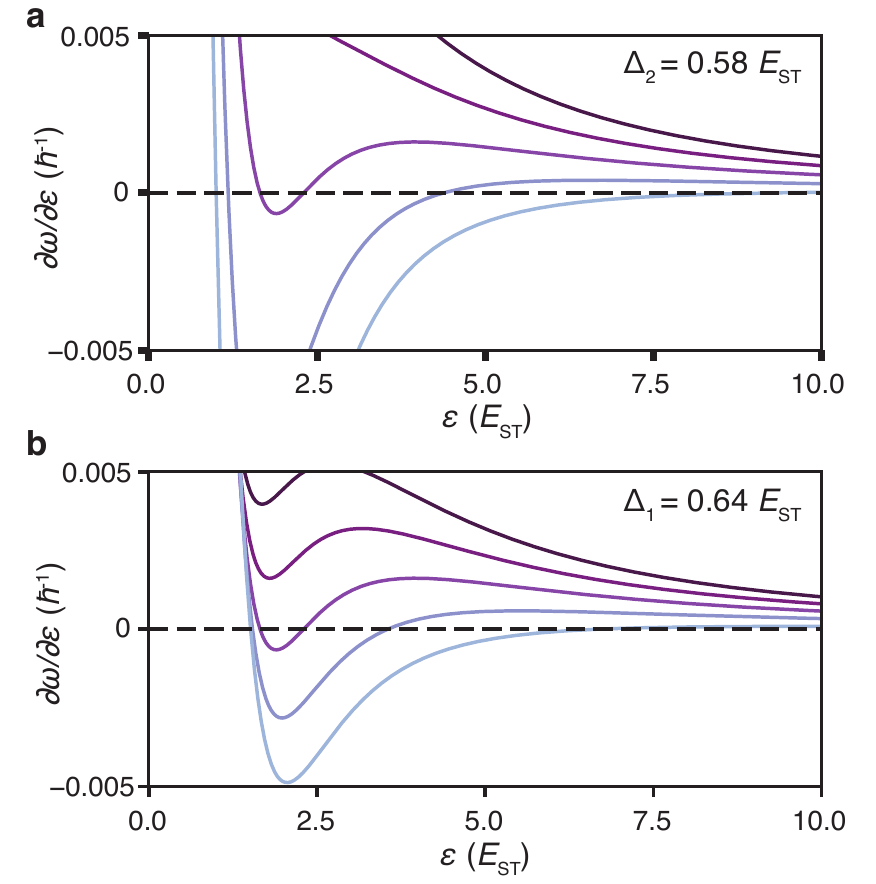} 
	\caption{{\bf Adjusting tunnel couplings to optimize single qubit dispersions.} 
The derivative of the QDHQ frequency $\omega$ with respect to the detuning $\varepsilon$, plotted as a function of $\varepsilon$. 
{\bf a}, Here we consider a fixed value of $\Delta_2 = 0.58 E_\text{ST}$ and several values of $\Delta_1$. 
From top to bottom, in units of $E_\text{ST}$: $\Delta_1= $0.6, 0.62, 0.64, 0.66, and 0.68. 
{\bf b}, Here we consider a fixed value of $\Delta_1 = 0.64 E_\text{ST}$ and several values of $\Delta_2$.
From top to bottom, in units of $E_\text{ST}$: $\Delta_2=$ 0.54, 0.56, 0.58, 0.6, and 0.62. 
The combination of $\Delta_1 = 0.64 E_\text{ST}$ and $\Delta_2 = 0.58 E_\text{ST}$ yields a dispersion for which $\hbar|\partial \omega / \partial \varepsilon|< 0.002$ for all detunings $\varepsilon >1.48 E_\text{ST}$. 
\label{figsupp1}}
\end{figure}

For the QDHQ, the qubit energy dispersion can be made extremely flat at large detunings by choosing the special values $\Delta_1 = \Delta_2 = E_\text{ST}/\sqrt{2}$.
However, faster two-qubit gates can be achieved by working at somewhat lower detuning values, where the qubits have large dipole moments. 
In Fig.~\ref{figsupp1} we plot $\partial \omega / \partial \varepsilon$ for a range of $\Delta_1$ and $\Delta_2$. 
Here we note that large dipole moments occur in the vicinity of the dips in the dispersion. 
The pulse sequences shown in Fig.~\ref{fig1}f of the main text require spending time at very large detuning values, and then transitioning adiabatically (i.e., slowly) to a region with such fast gates. 

To reduce the dephasing effects caused by charge noise, we therefore choose values of $\Delta_1$ and $\Delta_2$ that provide relatively flat dispersions over a wide range of epsilon.
In Fig.~\ref{figsupp1} this corresponds to tunnel couplings given by $\Delta_1 = 0.64 E_\text{ST}$ and $\Delta_2 = 0.58 E_\text{ST}$, and $\Delta_1/\Delta_2\simeq 1.103$.
For these values, the qubit can be operated near the charge transition point ($\varepsilon\simeq 0$), where gates are fast, while still maintaining long single-qubit coherence times.

\begin{widetext}
\section{Pulse sequences}
The simple detuning-only entangling pulse sequence considered in this work is shown in Fig.~1f of the main text.
It is parameterized by the four quantities $\varepsilon^{(1)}_{\text{wait}}$, $\varepsilon^{(2)}_{\text{wait}}$, $\tau_{\text{ramp}}$, and $\tau_{\text{wait}}$. 
Here we provide the explicit sequence used in our simulations, including the ramp functions:
\begin{equation}
\varepsilon^{(i)} = \left\{ \begin{array}{cc}
\varepsilon^{(i)}_{\text{init}} & t<0 , \\
\varepsilon^{(i)}_{\text{init}} + \left( \varepsilon^{(i)}_{\text{wait}}-\varepsilon^{(i)}_{\text{init}} \right)\sin^2\left( \frac{\pi t}{2\tau_{\text{ramp}}}\right) & 0<t<\tau_{\text{ramp}} , \\
 \varepsilon^{(i)}_{\text{wait}} & \tau_{\text{ramp}}<t<\tau_{\text{ramp}}+\tau_{\text{wait}} , \\
\varepsilon^{(i)}_{\text{init}} - \left( \varepsilon^{(i)}_{\text{wait}}-\varepsilon^{(i)}_{\text{init}} \right)\sin^2\left( \frac{\pi (t-\tau_{\text{ramp}}-\tau_{\text{wait}})}{2\tau_{\text{ramp}}}\right) & \tau_{\text{ramp}}+\tau_{\text{wait}}<t<2\tau_{\text{ramp}}+\tau_{\text{wait}} , \\
\varepsilon^{(i)}_{\text{init}} &  2\tau_{\text{ramp}}+\tau_{\text{wait}}<t .
\end{array}\right.
\label{pulseSeq}
\end{equation}
In our simulations we adopt the initial detuning values $\varepsilon^{(1)}_{\text{init}}=\varepsilon^{(2)}_{\text{init}}=500$ $\mu$eV, for which effective one- and two-qubit couplings are negligible.
These values are then varied during the optimization procedure.

To explore the possibility of a dynamical sweet spot, we also consider the tunnel-coupling pulse sequence shown in Fig.~3b of the main text, which is parameterized by the variables listed in Table~SI.
Again we provide the explicit sequence used in our simulations, including the ramp functions:
\begin{figure*}[h]
	\begin{equation}
	\Delta(t)= \left\{ \begin{array}{cc}
	\Delta_{\text{init}} & t \leq t_1 \\
	\Delta_{\text{init}} + \left( \Delta^\text{I}_\text{wait}-\Delta_{\text{init}} \right)\sin^2\left( \frac{\pi (t-t_1)}{\tau^\text{I}_\text{ramp}}\right) & t_1 \leq t<t_2 \\
	\Delta^\text{I}_\text{wait} & t_2 \leq t<t_3\\
    \Delta^\text{I}_\text{wait} + \left( \Delta^\text{II}_\text{wait}-\Delta^\text{I}_\text{wait}\right)\sin^2\left( \frac{\pi (t-t_3)}{\tau^\text{II}_\text{ramp}}\right) & t_3 \leq t<t_4 \\
	\Delta^\text{II}_\text{wait} &  t_4 \leq t<t_5,
	\end{array}\right.
	\label{pulseSeqTunnel}
	\end{equation}
\end{figure*}

\noindent
where $t_1 = \tau_{\text{wait}}^0$, $t_2 = t_1 +\tau^\text{I}_\text{ramp}$, $t_3 = t_2 + \tau_{\text{wait}}^\text{I}$, $t_4 = t_3 + \tau^\text{II}_\text{ramp}$, and $t_5 = t_4 + \tau_{\text{wait}}^\text{II}/2$. For $t>t_5$,
 \begin{equation}
 \Delta(t) = \Delta(t_5-t).
 \end{equation}
For brevity here, we have dropped the superscripts and subscripts on the tunnel coupling parameters.
In this case, we adopt the initial values $\Delta^{(1)}_{1,\text{init}} = 33.28$ $\mu$eV, $\Delta^{(1)}_{2,\text{init}} = 30.16$ $\mu$eV, $\Delta^{(2)}_{1,\text{init}} = 28.8$ $\mu$eV, and $\Delta^{(2)}_{2,\text{init}} = 26.1$ $\mu$eV, which were chosen according to the considerations of Section S2, above. 
For this sequence, we also assumed the fixed values $\varepsilon^{(1)}_{\text{wait}} = 90$ $\mu$eV and $\varepsilon^{(2)}_{\text{wait}} = 110$ $\mu$eV, as discussed in the main text.
The optimized values obtained for the other parameters in the sequence are listed in Table~S1.

\begin{table*}[h]
	\centering
	\begin{tabular}{|c||c|c|c|c|c|c|c|c||c|c|}
		\hline
		&       & $\tau^0_\text{wait}$ (ns) & $\tau^\text{I}_\text{ramp}$ (ns) & $\tau^\text{I}_\text{wait}$ (ns) & $\tau^\text{II}_\text{ramp}$ (ns) & $\tau^\text{II}_\text{wait}$ (ns) & $\Delta^\text{I}_\text{wait}$ ($\mu$eV)  & $\Delta^\text{II}_\text{wait}$ ($\mu$eV) &   $\tau_\text{ramp}$ (ns) &  $\tau_\text{wait}$ (ns)   \\
		\hline
		\multirow{2}{*}{\begin{tabular}[c]{@{}c@{}}$\Delta_1^{(i)}/\Delta_2^{(i)}$ \\ =1.1034\end{tabular}}  &   $\Delta_1^{(1)}$    &   1.10     &    0.45     &    0.00     &     1.81     &   2.10       &      33.263      &     32.752    & \multirow{2}{*}{\begin{tabular}[c]{@{}c@{}} 2.847 \end{tabular}}   &    \multirow{2}{*}{\begin{tabular}[c]{@{}c@{}} 2.587 \end{tabular}}\\
		&   $\Delta_1^{(2)}$    &   1.14      &    0.31     &   1.40      &   0.76       &    1.83      &      35.245      &      28.9555     &    &      \\
		\hline
		\multirow{4}{*}{\begin{tabular}[c]{@{}c@{}}$\Delta_1^{(i)}/\Delta_2^{(i)}$ \\ free\end{tabular}}   &   $\Delta_1^{(1)}$     &   0.31      &     1.09    &    1.27     &     2.21     &   1.01       &     11.200       &      46.332      &   \multirow{4}{*}{\begin{tabular}[c]{@{}c@{}} 4.185 \end{tabular}} &    \multirow{4}{*}{\begin{tabular}[c]{@{}c@{}} 1.705 \end{tabular}}    \\
		&    $\Delta_2^{(1)}$    &    0.74    &   0.58      &   0.38      &    1.40      &    2.79      &    3.977        &      31.341     &    &      \\
		&    $\Delta_1^{(2)}$     &   1.06    &  2.01   &  0.00      &   0.69      &  2.21     &     20.974    &    20.828    &    &    \\
		&    $\Delta_2^{(2)}$    &   0.25     &    1.28     &     1.03    &    2.46      &    0.86      &      34.976      &       67.408     &    &     \\
		\hline           
	\end{tabular}
	\caption{Parameters used in the tunnel coupling pulse sequence shown in Fig.~3b of the main text, and accompanying detuning pulse sequence shown in Fig.~1f, obtained using the optimization procedure described in the main text.
	The specific ratio, $\Delta_1^{(i)}/\Delta_2^{(i)}=1.1034$ used in the first set of solutions is consistent with the discussion in Section~S2, above.}
	\label{table1}
\end{table*}
\end{widetext}

\section{Optimization of detuning pulse parameters}
The simple two-qubit detuning-only pulse sequence used in this work is shown in Fig.~\ref{fig1}f of the main text, and is completely defined by the parameters $\varepsilon^{(1)}_\text{wait}$, $\varepsilon^{(2)}_\text{wait}$, $\tau_\text{ramp}$, and $\tau_\text{wait}$, as described in Section~S3, above.
In the current section, we explain how these parameters are chosen in our analysis, while constraining the total gate time $2\tau_\text{ramp}+\tau_\text{wait}<50$~ns.
The procedure is summarized as follows.

\begin{figure}[!tp]
	\includegraphics[width=1.0 \linewidth]{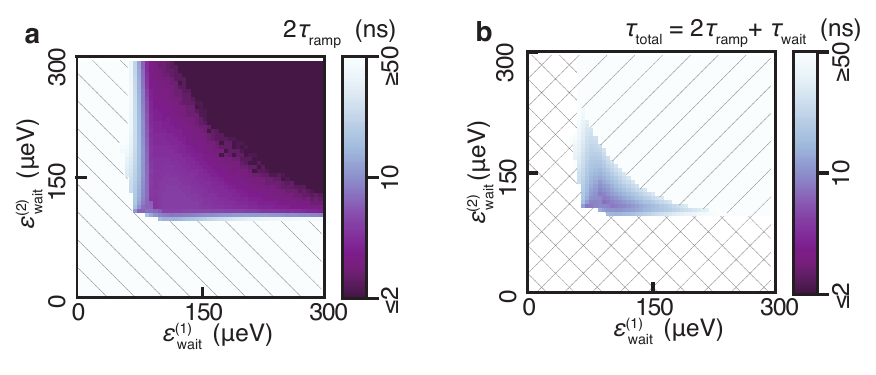} 
\caption{{\bf Key steps in the detuning-pulse optimization procedure, as described in Section~S4.} 
{\bf a}, Step 2: determine $\tau_\text{ramp}$.
We simulate the ramp portion of the pulse sequence in the absence of noise and compute the leakage fidelity $\mathcal{F}_\text{leak}$ that quantifies the adiabaticity of the pulse, defined as the probability that a system initialized into a two-qubit logical state remains in that state after the pulse is applied.
The plot shows the shortest values of $\tau_\text{ramp}$ giving $\mathcal{F}_\text{leak}>99.9$\%.
Results with $2\tau_\text{ramp}>50$~ns are omitted from the rest of the analysis because they are much slower than single-qubit gates (cross-hatched region).
{\bf b}, Step 4: determine $\tau_\text{wait}$.
We compute the noise-averaged fidelity $\langle\mathcal{F}\rangle$ of a full CZ pulse sequence, modulo single-qubit gate operations.
The plot shows the value of $\tau_\text{wait}$ that maximizes $\langle\mathcal{F}\rangle$.
Solutions with $\tau_\text{total}=2\tau_\text{ramp}+\tau_\text{wait}>50$~ns are now omitted.
\label{supp_opt_taus}}
\end{figure}

\begin{enumerate}[wide, labelwidth=!, labelindent=0pt, itemsep=1mm]

\item {\bf Choose specific values for $\bm{\varepsilon^{(1)}}_\text{wait}$ and $\bm{\varepsilon^{(2)}}_\text{wait}$.}

\item {\bf Determine $\bm{\tau}_\text{ramp}$.}
Consider a ``ramp-only" detuning sequence with $\tau_\text{wait}=0$. 
Non-adiabatic effects such as leakage occur \emph{only} during these ramp steps, and we determine $\tau_\text{ramp}$ by ensuring that it satisfies a ``fast-adiabatic" criterion, defined as follows.
We first define the leakage fidelity as $\mathcal{F}_\text{leak}=\frac{1}{4}\sum |\bra{ij}U_{2q}\ket{ij}|^2$, where the sum is taken over the logical states, $(i,j)=(0,1)$, defined at time $t=0$ in the pulse sequence, and $U_{2q}$ is the 9D unitary operator derived from $\mathcal{H}_{2q}$, and computed here in the absence of charge noise.
(Note that $\mathcal{F}_\text{leak}$ is defined similarly to $\mathcal{I}_\text{na}$ in Methods, except that here, the logical states on either side of the matrix element are evaluated at the initial time, and are the same as the logical states at the final time of the pulse sequence.)
We then choose a $\tau_\text{ramp}$ that corresponds to the shortest ramp time giving $\mathcal{F}_\text{leak}>99.9$\%.
Finally, we omit any solutions with $2\tau_\text{ramp}>50$~ns from the rest of the analysis; these correspond to the cross-hatched regions of Fig.~\ref{supp_opt_taus}a.
As discussed in the main text, the fidelity of the results shown in Fig.~3c is limited by charge noise, not leakage, indicating that the particular choice of $\mathcal{F}_\text{leak}>99.9$\% does not affect our final results.

\item {\bf Determine the angles $\bm{\phi_1}$ and $\bm{\phi_2}$,} defined such that $U_\text{4D}\simeq U_\text{ideal}$, where $U_\text{ideal}=Z^{(1)}(\phi_1)Z^{(2)}(\phi_2)\, \text{CZ}$.
Here, $U_\text{4D}$ is the projection of $U_{2q}$ onto the 4D logical subspace; in this step, $U_{2q}$ is computed as a function of $\tau_\text{wait}$ in the absence of charge noise (but including leakage), using the values of $\varepsilon^{(1)}_\text{wait}$, $\varepsilon^{(2)}_\text{wait}$, and $\tau_\text{ramp}$ chosen above.
CZ is defined as $\text{diag}\{1,1,1,-1\}$.
The explicit method used to determine $\phi_1$ and $\phi_2$ as a function of $\tau_\text{wait}$ is given as follows.
First, we adjust the overall phase of $U_{2q}$ such that $\bra{00}U_{2q}\ket{00}$ is real and approximately equal to 1.
Then we define $\phi_1=\text{angle}[\bra{01}U_{2q}\ket{01}]+\pi/2$ and $\phi_2=\text{angle}[\bra{10}U_{2q}\ket{10}]+\pi/2$, where the function $\text{angle}[u]$ gives the complex phase of $u$.

\item {\bf Determine $\bm{\tau}_\text{wait}$.}
Recompute $U_{2q}$ as a function of $\tau_\text{wait}$, now including charge noise.
Compute the process fidelity $\mathcal{F}$, as described in Section~S5 below, for each value of $\tau_\text{wait}$, where the actual $\chi$ matrix is obtained from $U_{2q}$, while the ideal $\chi$ matrix is derived from $U_\text{ideal}$.
Perform an average of $\mathcal{F}$ over charge noise configurations, as described in Methods.
Maximize this $\langle\mathcal{F}\rangle$ with respect to $\tau_\text{wait}$, omitting any results for which $\tau_\text{total}=2\tau_\text{ramp}+\tau_\text{wait}>50$~ns.

\item Finally, {\bf choose the optimal values of $\bm{\varepsilon^{(1)}}_\text{wait}$ and $\bm{\varepsilon^{(2)}}_\text{wait}$}, as described in the main text, by determining the maximum fidelity shown in Fig.~2.

\end{enumerate}

\section{Process Fidelity}
In Figs.~2 and 3 of the main text, we report fidelities that are averaged over a noise distribution.
To compute the fidelity for a given instance of noise, we first solve Eq.~\eqref{unitary} for the appropriate pulse sequence to obtain the corresponding unitary operator $U_{2q}$. 
The process fidelity is defined as $\mathcal{F} = \text{Tr}(\chi_{\text{ideal}}\chi)$, where $\chi$ is the actual process matrix, including noise and leakage effects, and $\chi_\text{ideal}$ is the ideal process matrix derived from $U_\text{ideal}$.
In this case, we use the 4D $U_\text{ideal}$ defined above, in Section~S4, which is then embedded in the full 9D Hilbert space.

\begin{figure}[tb]
	\includegraphics[width=1.0 \linewidth]{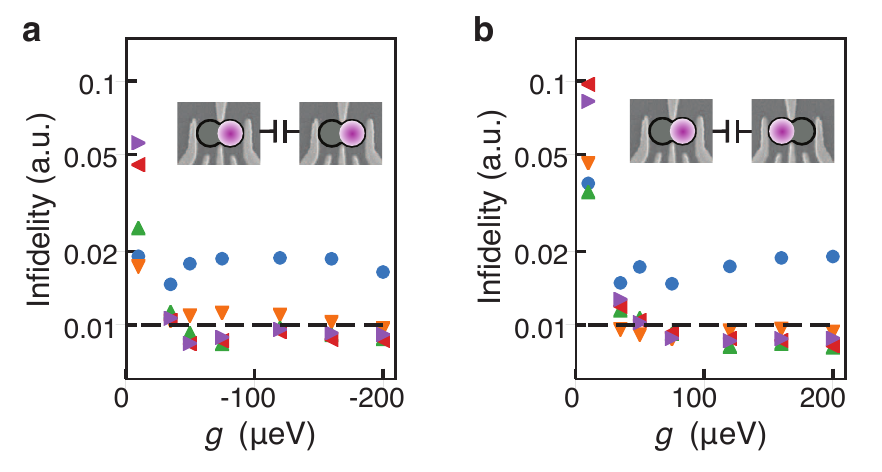} 
	\caption{{\bf Optimized process infidelities of adiabatic CZ gates as a function of $\Delta E_\text{ST}$ and $g$.} 
Results are obtained using the procedure described in Section~S4.
{\bf a}, $g<0$.
{\bf b}, $g>0$.
In both cases, the markers correspond to $\Delta E_\text{ST} = 1$~$\mu$eV (blue circles), $\Delta E_\text{ST} = 3$~$\mu$eV (orange downwards-facing triangles), $\Delta E_\text{ST} = 5$~$\mu$eV (green upwards-facing triangles), $\Delta E_\text{ST} = 7$~$\mu$eV (red left-facing triangles), and $\Delta E_\text{ST} = 9$~$\mu$eV (purple right-facing triangles). The threshold for acceptable infidelity is chosen to be 1\% (black dashed lines). 
Insets: {\bf a}, Aligned dipoles ($g<0$); {\bf b}, Anti-aligned dipoles ($g>0$).
\label{figsupp3}}
\end{figure}

We obtain the process matrix $\chi$ from $U_{2q}$ using the Choi-Jamiolkowski isomorphism, \cite{PhysRevA.71.062310} in which the process matrix is given by $\chi = d\rho$, where $d$ is the dimensionality of the system (in this case, $d=9$), and $\rho$ is given by
\begin{equation}
\rho = \left[I\otimes U_{2q}\right]\left(\ket{\Phi}\bra{\Phi}\right).
\end{equation}
Here, $I$ is the identity matrix of the 9D Hilbert space and the Jamiolkowski state $\ket{\Phi}$ is defined as
\begin{equation}
\ket{\Phi} = \frac{1}{2}\sum \ket{ij}\ket{ij},
\end{equation}
where the sum is taken over the logical eigenstates $(i,j)=(0,1)$ of the two-qubit Hamiltonian described in Eq.~\eqref{two_qubit} of the main text. 
Note that the Jamiolkowski state only includes four states despite the full system having nine states.

\section{Dependence of the fidelity on $g$ and $\Delta E_\text{ST}$}
In the CZ gate analyses presented in Figs.~2 and 3 of the main text, $g$ was chosen to be $75$ $\mu$eV, as this was an experimentally measured value. \cite{Ward:2016aa} Additionally, the singlet-triplet splittings $E_\text{ST}^{(1)}$ and $E_\text{ST}^{(2)}$ were chosen to be 52~$\mu$eV and~47 $\mu$eV, respectively, inspired by experimentally measured values. \cite{Thorgrimsson:2017aa} However, both $g$ and the singlet-triplet splittings will vary from experiment to experiment; indeed, capacitive couplings as large as $200$ $\mu$eV have been measured. \cite{2016PhRvP...6e4013Z} 
In this section, we determine whether either $g$ or the difference between the singlet-triplet splittings, $\Delta E_\text{ST}= E_\text{ST}^{(1)}-E_\text{ST}^{(2)}$, can be used to further optimize the fidelity, finding no significant improvements.

We first note that it is possible to change the sign of the Coulomb interaction $g$ by reversing the  alignment of the charge dipole of one of the qubits (say, $i$), which amounts to changing the sign of $\varepsilon^{(i)}$.
Here, we adopt the convention in Eq.~\eqref{two_qubit} of the main text that $g<0$ corresponds to the dipoles being aligned in the limit of large detunings, as indicated in the inset of Fig.~\ref{figsupp3}a, while $g>0$ corresponds to the dipoles being anti-aligned, as indicated in the inset of Fig.~\ref{figsupp3}b.
Clearly the sign of $g$ also affects the qubits' tendency to align or anti-align as the detunings are varied, and we therefore expect our results to depend on this sign.

In Fig.~\ref{figsupp3}, we plot the noise-averaged infidelities obtained for a range of $g$ (both positive and negaive values) and $\Delta E_\text{ST}$, assuming a constant level of quasistatic charge noise, $\sigma_\varepsilon = 1$ $\mu$eV.
The results are obtained using the procedure described in Section S4 to obtain the absolute minimum infidelity for the detuning-only pulse sequence.
We find that the infidelities generally fall below a threshold criterion of 0.01 (i.e., 1\%), except when $-35\leq g\leq 50$ $\mu$eV, or when $\Delta E_\text{ST}\leq 3$ $\mu$eV (with $g<0$), or $\Delta E_\text{ST}\leq 1$ $\mu$eV (with $g>0$).

 \begin{figure}[tb]
	\includegraphics[width=1.0 \linewidth]{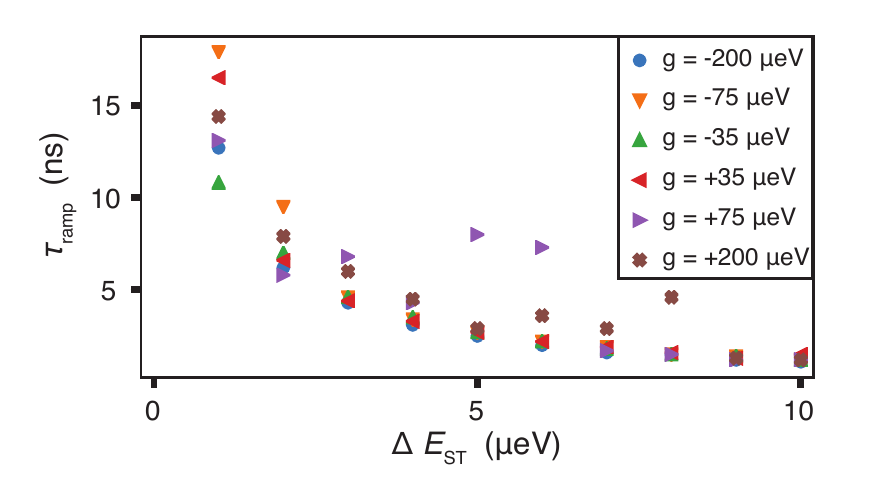} 
	\caption{{\bf Adiabatic ramp time diverges for small $\Delta E_\text{ST}$.} 
Ramp time $\tau_{\text{ramp}}$ obtained by optimizing the process fidelity of the detuning pulse sequence is plotted as a function of $g$ and $\Delta E_\text{ST}= E_\text{ST}^{(1)}-E_\text{ST}^{(2)}$. 
When $\Delta E_\text{ST}\rightarrow 0$, the resulting energy level degeneracy causes $\tau_\text{ramp}$ to diverge, to preserve adiabaticity, thus increasing the exposure to charge noise and the infidelity. \label{figsupp2}}
\end{figure}

We can understand the behaviors observed in Fig.~\ref{figsupp3} as follows.
The infidelity decreases with $|g|$ because the entanglement frequency $\omega_{zz}$ is roughly proportional to $g$, as observed in Eq~\eqref{coupling}. 
When $g$ is small, the entangling gate speed is therefore slow.
This must be compensated by reducing $\varepsilon^{(1)}$ and $\varepsilon^{(2)}$; however this also increases the susceptibility to charge noise, and  the infidelity.

The dependence on $\Delta E_\text{ST}$ in Fig.~\ref{figsupp3} can be understood by noting that the limit $\Delta E_\text{ST}\rightarrow 0$ corresponds to the degeneracy of logical states $\ket{01}$ and $\ket{10}$ in the limit of large detunings.
Degenerate energy levels cause problems for adiabatic operation, which can only be solved by reducing the ramp speed.
This is demonstrated in Fig.~\ref{figsupp2} where we plot the optimized value of $\tau_\text{ramp}$ as a function of $g$ and $\Delta E_\text{ST}$.
Here we observe little dependence on $g$, but for small $\Delta E_\text{ST}$, $\tau_\text{ramp}$ and therefore $\tau_\text{total}$ increase significantly.
The longer gates are more exposed to charge noise, resulting in lower fidelity.
On the other hand, for $\Delta E_\text{ST}>3$~$\mu$eV, the total gate time is dominated by the waiting time $\tau_\text{wait}$, so the further reduction of $\tau_\text{ramp}$ has a marginal effect on the infidelity.

 \begin{figure}[t]
	\includegraphics[width=0.95 \linewidth]{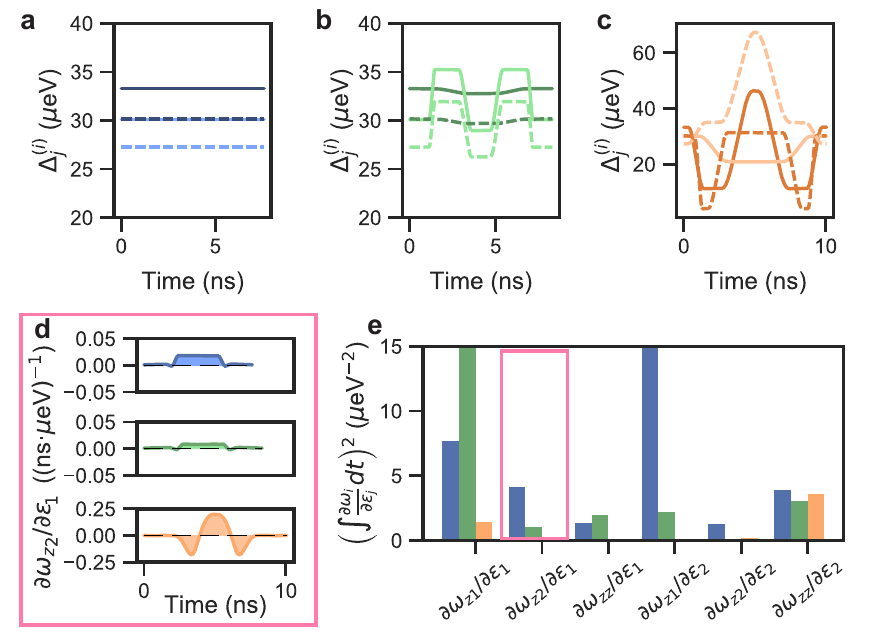} 
	\caption{{\bf Pulse sequences involving both the tunnel coupling and detuning parameters are able to improve the fidelity of a CZ gate.} 
The tunnel coupling pulses shown in Fig.~\ref{fig3} of the main text improve the process fidelity by lowering the time-averaged value of the derivatives $\partial \omega_j/\partial \varepsilon_i$. 
As described in the main text, optimized results are obtained for three different tunnel coupling pulse sequences when
{\bf a}, tunnel couplings are held fixed for the duration of the sequence,
{\bf b}, only the ratios $\Delta_1^{(1)}/\Delta_2^{(1)}=\Delta_1^{(2)}/\Delta_2^{(2)}$ are held fixed throughout the sequence, and
{\bf c}, no constraints are placed on the tunnel coupling sequence parameters.
Here, the dark solid lines correspond to $\Delta^{(1)}_1$, the dark dashed lines correspond to $\Delta^{(1)}_2$, the light solid lines correspond to $\Delta^{(2)}_1$, and light dashed lines correspond to $\Delta^{(2)}_2$. 
The color codings are the same as in Fig.~3c of the main text.
{\bf d}, The derivative $\partial\omega_{z2}/\partial{\varepsilon_1}$ is plotted as a function of time, showing a change of sign due to the application of the tunnel coupling pulse sequence.
{\bf e}, The squared, time-averaged derivatives are shown for all the different qubit frequencies.
For a perfect dynamical sweet spot, these integrals would all vanish. 
Here, the pink box corresponds to the results shown in {\bf d}.
		\label{Supp_fig4}}
\end{figure}

\section{Approximate formula for the Infidelity}
In this section, we derive an approximate analytical expression for the charge-noise induced infidelity, $\mathcal{I}_\text{cn}$, to more efficiently identify tunnel coupling pulse sequences that improve the gate fidelity. 

In the absence of non-adiabatic processes, we can evaluate the effective two-qubit Hamiltonian $\mathcal{H}_{2q}$ in its adiabatic basis, giving Eq.~(3) of the main text.
Since this Hamiltonian is strictly diagonal, it is trivial to compute the resulting unitary operator for the logical subspace: 
\begin{align*}
U_\text{4D} = \text{diag}(&\exp{(i(-\theta_{z1}-\theta_{z2}+\theta_{zz})/2)},\\
&\exp{(i(\theta_{z1}-\theta_{z2}-\theta_{zz})/2)},\\
&\exp{(i(-\theta_{z1}+\theta_{z2}-\theta_{zz})/2)},\\
&\exp{(i(\theta_{z1}+\theta_{z2}+\theta_{zz})/2)}),
\end{align*} 
where $\theta_i=\int \omega_i dt$.
Quasistatic charge noise causes the phases to evolve with errors defined as $\Delta\theta_i = \theta_i-\theta_i^\text{ideal}$.
However, the resulting time evolution is unitary, and the methods of Supplementary Section~S5 easily give an expression for the process infidelity:
\begin{gather}
\mathcal{I} = 1-\frac{1}{8}[2+\cos(\Delta\theta_{z1}+\Delta\theta_{z2})+\cos(\Delta\theta_{z1}-\Delta\theta_{z2})\nonumber\\
+2\left[\cos(\Delta\theta_{z1})+\cos(\Delta\theta_{z2})\right]\cos(\Delta\theta_{zz})]\label{inf_exp} .
\end{gather}
Again for quasistatic charge noise, we can approximate
\begin{equation}
\Delta\theta_i \approx \delta\varepsilon^{(1)}\int\frac{\partial \omega_i}{\partial \varepsilon^{(1)}} dt + \delta\varepsilon^{(2)}\int\frac{\partial \omega_i}{\partial \varepsilon^{(2)}} dt,
\end{equation}
where $\delta\varepsilon^{(j)}$ is the noise on $\varepsilon^{(j)}$, which is assumed to be constant over the duration of the pulse sequence.
Substituting these definitions into Eq.~\eqref{inf_exp}, expanding in small $\Delta\theta_i $, and averaging over the noise distribution as described in Methods yields Eq~\eqref{quadDerivatives} of the main text.

\section{DSS Analysis for the tunnel coupling pulse sequence}

In Fig.~3 of the main text, we showed optimized infidelity results for three different tunnel coupling pulse sequences.
The optimized pulse sequences are shown in Fig.~\ref{Supp_fig4}a-c with the same color coding as Fig.~3. 
In Fig.~\ref{Supp_fig4}a the tunnel couplings are held fixed for the duration of the detuning pulse. Figure~\ref{Supp_fig4}b shows the non-constant tunnel coupling pulse sequence obtained under the constraint that the ratios $\Delta_1^{(1)}/\Delta_2^{(1)}=\Delta_1^{(2)}/\Delta_2^{(2)}=1.1034$ are held fixed throughout the pulse sequence. 
Figure~\ref{Supp_fig4}c shows the pulse sequence obtained when the tunnel couplings are allowed to vary without constraint. 

A DSS is formed when the time-averaged derivatives of the qubit frequencies go to zero, as described below Eq.~(4) of the main text.
We plot these time averages in Fig.~\ref{Supp_fig4}e for each of the different pulse sequences, with details of the time dependence for one of the qubit frequencies shown in Fig.~\ref{Supp_fig4}d.
As seen here, certain choices for the pulse sequence cause the derivatives to change sign as a function of time, leading to an overall suppression of the time average and the infidelity.
We note that the effect is especially pronounced for the pulse sequence with the largest number of tuning parameters.
However, we also note that the derivative of the $\omega_{zz}$ qubit frequency is particularly difficult to suppress.

\end{document}